\documentclass[article,twocolumn,showpacs,aps,prb,floatfix]{revtex4-1}
\usepackage{graphicx,subfigure,epsfig,verbatim,psfrag,amsmath,amssymb}
\usepackage{indentfirst}
\usepackage{textcomp}
\usepackage{latexsym}
\usepackage{amssymb}
\usepackage{amsmath}
\usepackage{lipsum}
\usepackage{epstopdf}
\usepackage{float}
\usepackage{multirow}
\usepackage{soul}
\usepackage[dvipsnames]{xcolor}
\newcommand\numberthis{\addtocounter{equation}{1}\tag{\theequation}}
\makeatletter
\begin{document}
\title{Interaction induced edge states in HgTe/CdTe  Quantum Well under magnetic field}
\author{Zewei Chen}
\author{Tai Kai Ng}
\email{phtai@ust.hk}
\affiliation{Physics Department,The Hong Kong University of Science and Technology, Hong Kong}

\begin{abstract}
In this paper, we study doped HgTe/CdTe quantum well with Hubbard-type interaction under perpendicular magnetic field using a lattice Bernevig-Hughes-Zhang (BHZ) model with a bulk inversion asymmetry (BIA) term. We show that the BIA term is strongly enhanced by interaction around the region when the band inversion of the topological insulator is destroyed by a magnetic field. The enhanced BIA term creates edge-like electronic states which can explain the experimentally discovered edge conductance in doped HgTe/CdTe quantum well at similar magnetic field regime.
\end{abstract}
\maketitle

\section{Introduction}  2D topological insulators have been extensively studied both theoretically and experimentally since its discovery\cite{Bernevig15122006,type2qsh,RevModPhys.82.3045,RevModPhys.83.1057,1002.2904,QSHETAE,QSHI_HgTe,Single_valley,Spin_polarization}. The non-trivial topology and the helical edge states are protected by the time reversal symmetry (TRS)\cite{topologicalinvariant}. 
The  detailed behavior of HgTe/CdTe quantum well under perpendicular magnetic field has been studied both experimentally\cite{Konig766} and theoretically\cite{PhysRevB.86.075418,effectofMFTI,BallisticQSHS}. It is believed that a transition from quantum spin hall (QSH) state  to integer quantum Hall (IQH) state occurs when the magnetic field is strong enough.  The Landau level fan charts (LLFC) shows a crossing at a critical magnetic field $B_c$ where the band inversion disappears. The helical edge state is destroyed around the transition regime and (chiral) edge states emerge when the system transits into the IQH state. When a bulk inversion asymmetry term is included,  the electron- and hole- bands hybridize and crossing is avoided. These results have been confirmed by magnetospectroscopy studies in HgTe/CdTe quantum well\cite{orlita2011fine,PhysRevB.86.205420}.

Edge transports under perpendicular magnetic field has been studied by Du's group in InAs/GaSb quantum well\cite{PhysRevLett.114.096802} at magnetic field range believed to be below $B_c$\cite{PhysRevB.90.115305}. Shen's group measured the local conductance under a perpendicular magnetic field in doped HgTe/CdTe quantum well \cite{unexpectededge} and found that the edge conductance persists under strong magnetic field up to 9T, much larger than the expected critical field $B_c$ but is still not strong enough to reach the IQH regime - the electron/hole filling factor in the experiment is still too small to fill the zeroth LL. Furthermore, the edge conductance exists only when it is electron-like gated indicating the importance of particle-hole asymmetry.  The non-interacting BHZ model is not able to explain these results and suggests that electron interaction may be important to understand the  HgTe/CdTe system\cite{unexpectededge}.

In this  paper we study the interaction effect in doped HgTe/CdTe quantum well {\em via} a modified lattice BHZ model that takes into account the Bulk-Inversion Asymmetry (BIA) term  and with Hubbard-type on-site interaction. The BIA term is found to be small in band structure calculations and is usually neglected. We find that BIA is enhanced by the combined effect of interaction and magnetic field in a self-consistent mean field theory.  The enhanced BIA term gives rise to edge-like states around the region when the band inversion is destroyed by a magnetic field and can explain the experimental result on the HgTe/CdTe quantum well by Shen's group\cite{unexpectededge}.

\section{Model}
We consider the BHZ model with BIA asymmetry term and Hubbard-type on-site interaction on a square lattice with two  orbital $\{|E\sigma\rangle,|H\sigma \rangle \}$ per site. The BIA term is allowed because HgTe/CdTe has a Zinc-blende structure which breaks bulk inversion symmetry\cite{QSHETAE}. We also apply a magnetic field perpendicular to the lattice plane.  The system is described by the Hamiltonian $H=H_{BHZ}+H_{BIA}+H_{z}+H_{U}$, where $H_{BHZ}=T+H_0$ is the (lattice) BHZ model with
\begin{subequations}
\begin{align}
H_{0}&=\sum_{i,\sigma} \left(\varepsilon_E C_{i,E,\sigma}^{\dag}  C_{i,E,\sigma} +\varepsilon_H C_{i,H,\sigma}^{\dag}C_{i,H,\sigma}\right),
\end{align}
where $\varepsilon_{\tau}$ is the on-site energy for $\tau$ orbital, $C^{\dagger}(C)_{i,\tau,\sigma}$ creates/annihilates a $\tau$-orbit ($\tau$=E,H) electron with spin $\sigma=\uparrow,\downarrow$  on site $i$ and
\begin{align}
T&=\sum_{\langle i,j \rangle,\sigma}\left(t_E C_{i,E,\sigma}^{\dag}C_{j,E,\sigma}  + t_H C_{i,H,\sigma}^{\dag}C_{j,H,\sigma}\right) \\
&+\sum_{i,\sigma}t_{EH}\left[s(i C_{i,E,\sigma}^{\dag}C_{i+\hat{x},H,\sigma} -i C_{i,E,\sigma}^{\dag}C_{i-\hat{x},H,\sigma})\right. \\
&+\left.( C_{i,E,\sigma}^{\dag}C_{i+\hat{y},H,\sigma} -C_{i,E,\sigma}^{\dag}C_{i-\hat{y},H,\sigma})\right]+\text{H.c.}
\end{align}
 describes electron hopping between nearest neighbor (NN) lattice sites $<i,j>$ where $t_{\tau}, t_{EH}$ denotes intra-orbital and inter-orbital hopping, respectively. $s=+(-)1$ for $\sigma=\uparrow(\downarrow)$. H.c. denotes the hermitian conjugate.
\begin{align}
H_{BIA}=-\Delta_0\sum_{i}( C_{i,E,\uparrow}^{\dag}C_{i,H,\downarrow}- C_{i,H,\uparrow}^{\dag}C_{i,E,\downarrow})+\text{H.c.}
\end{align}
is the BIA term where $\Delta_0\sim1.5-2meV$ and
\begin{align}
H_{z}=\sum_{i,\tau,\sigma} s g_{\tau}\mu_B B_z C_{i,\tau,\sigma}^{\dag}C_{i,\tau,\sigma}
\end{align}
\end{subequations}
 is the Zeeman energy. $\mu_B$ is the Born  magneton and $g_{\tau}$ is the g-factors for $\tau$-orbit. $B_z$ is the magnetic field strength. The orbital magnetic field effect is included by Peierls substitution, $t_{\tau}\to t_{\tau} \exp\left(i 2\pi (j-1) \Phi / \Phi_0 \right)$ with gauge field $\textbf{A}=-B_z y \hat{x}$ (Landau gauge). $\Phi=B_z a^2$ is the magnetic flux passes through a lattice cell and $\Phi_0=h/e$ is the magnetic flux quantum.

\begin{align}
H_{U}=\sum_{i;\tau=E,H} U_{\tau} n_{i,\tau,\uparrow} n_{i,\tau,\downarrow}+\sum_{i;\sigma,\sigma'} U_{EH} n_{i,E,\sigma} n_{i,H,\sigma'}
\end{align}
where $U_{\tau} (\tau=E,H), U_{EH}>0$ describe intra- and inter- orbital repulsive interaction between electrons, respectively, $n_{i,\tau,\sigma}=C_{i,\tau,\sigma}^{\dag} C_{i,\tau,\sigma}$.

We shall treat the interaction term in a mean-field theory where
\begin{align}
\begin{split}
&n_{i,\tau,\sigma}n_{i,\tau',\sigma'}\approx \langle n_{i,\tau,\sigma} \rangle n_{i,\tau',\sigma'} + \langle n_{i,\tau',\sigma'} \rangle n_{i,\tau,\sigma}\\
 &- \langle n_{i,\tau,\sigma} \rangle  \langle n_{i,\tau',\sigma'} \rangle
\\
&-\left(\langle C_{i,\tau,\sigma}^{\dag}C_{i,\tau',\sigma'}\rangle C_{i,\tau',\sigma'}^{\dag}C_{i,\tau,\sigma}
+ \langle C_{i,\tau',\sigma'}^{\dag}C_{i,\tau,\sigma} \rangle C_{i,\tau,\sigma}^{\dag}C_{i,\tau',\sigma'}\right.\\
&\left.-\langle C_{i,\tau',\sigma'}^{\dag}C_{i,\tau,\sigma} \rangle \langle C_{i,\tau,\sigma}^{\dag}C_{i,\tau',\sigma'}\rangle\right) \delta_{\bar{\tau},\tau'}\delta_{\sigma,-\sigma'}
 \end{split} \nonumber
 \label{mean_field}
\end{align}
where $\bar{E}(\bar{H})=H(E)$ and $\langle...\rangle$ denotes ground state expectation value.  We note that the on-site hybridization term between the $E$ and $H$ orbital vanishes because of the opposite parity of the two orbital. The mean field Hamiltonian is therefore,
\begin{align}
\begin{split}
H_{MF}&=H_{BHZ}+ H_{BIA}+ H_{z}
\\&+\sum_{i,\sigma,\tau}(U_{\tau} \langle n_{i,\tau,-\sigma} \rangle +U_{EH} \langle n_{i,\bar{\tau}} \rangle) n_{i,\tau,\sigma}\\
&-U_{EH}(\Delta_1 C_{i,E,\uparrow}^{\dag}C_{i,H,\downarrow}-\Delta_{2} C_{i,H,\uparrow}^{\dag}C_{i,E,\downarrow}+\text{H.c.})
\end{split}
\end{align}
where  $\Delta_{1(2)}=+(-)\langle C_{i,H(E),\downarrow}^{\dag}C_{i,E(H),\uparrow}\rangle$ couples the spin up electron(hole) orbital to spin down hole(electron) orbital, respectively and $n_{i,\tau}=\sum_{\sigma}n_{i,\tau,\sigma}$.
We note that our mean-field theory allows an interaction-modified BIA term $\Delta_0\rightarrow \tilde{ \Delta}_{1(2)}=\Delta_0+ U_{EH}\Delta_{1(2)}$ and also possibility of magnetic phases with $\langle n_{i,\tau,\sigma} \rangle\neq\langle n_{i,\tau,-\sigma} \rangle$. The mean-field parameters and phase diagram are determined numerically in our study.

We consider the  half-filled BHZ model where the chemical potential is in the gap and the system is a topological insulator. To describe the experimental material\cite{unexpectededge},  we start with the parameters appropriate for the 7.5nm HgTe/CdTe quantum well with $\varepsilon_{E}=C+M-4(B+D)/a^2$,$\varepsilon_{H}=C-M+4(B-D)/a^2$, $t_{E}=(D+B)/a^2$,$t_{H}=(D-B)/a^2$,$t_{EH}=A/2a$ where  $C,M,B,D,A$ and $g_{|tau|}$ are the parameters in BHZ model determined in Ref.[\onlinecite{ReentrantTP}] (see Appendix A for details). We note however that the band-structure parameters can be changed quite significantly upon doping which is the case of the doped material HgTe/Hg$_{0.3}$Cd$_{0.7}$Te (7.0nm HgTe/CdTe quantum well)\cite{Bernevig15122006,QSHETAE} where the sign of $D/B$ is found to be inverted in the doped material, corresponding to changing the light-electron, heavy-hole bands into heavy-electron, light-hole  bands\cite{QSHETAE}. We believe that this is also happening in 7.5nm material for reason which will become clear later.  Therefore, we choose the parameters in our tight binding model to be: $t_{E}=-0.42 eV, t_{H}=3.32 eV, t_{EH}=0.275 eV,\varepsilon_{E}=1.67 eV, \varepsilon_{H}=-13.27 eV$, corresponding to changing $D\rightarrow-D$ in Ref.[\onlinecite{ReentrantTP}]. We also set $C=0$ in our calculation since it can be absorbed in the chemical potential.  The lattice constant $a$ is chosen to be 1nm. The phase diagram and mean-field parameters are studied under perpendicular magnetic field $B_z$ with these parameters for various values of $U_H, U_E$ and $U_{EH}$.  We have performed the calculation at $B_z=0$ and several values of $B_z\geq 3.5T$. We note that the magnetic unit cell becomes too large for numerical calculation for $B_z < 3.5T$.

\section{Results}

The mean field parameters are determined self-consistently. We first discuss the mean field phase diagram in absence of magnetic field. We find that the system is in the normal, non-magnetic state $\left( \langle n_{i,\tau,\sigma} \rangle=\langle n_{i,\tau,-\sigma} \rangle \right)$ for small $U_{E}, U_{H}$ and $U_{EH}$. For given $U_{EH}$ and $U_{E}$, the system transits from {\em paramagnetic }phase to {\em ferromagnetic }phase and then to {\em anti-ferromagnetic } phase as $U_{H}$ increases. The phase diagram can be understood by  comparing the model with the single band Hubbard model whose mean field phase diagram is well studied. We referred the readers to Appendix B for details.

\begin{figure}[tbh!]
\centering
\mbox{
\subfigure[\label{fig:delta1}]{\includegraphics[width=0.23\textwidth]{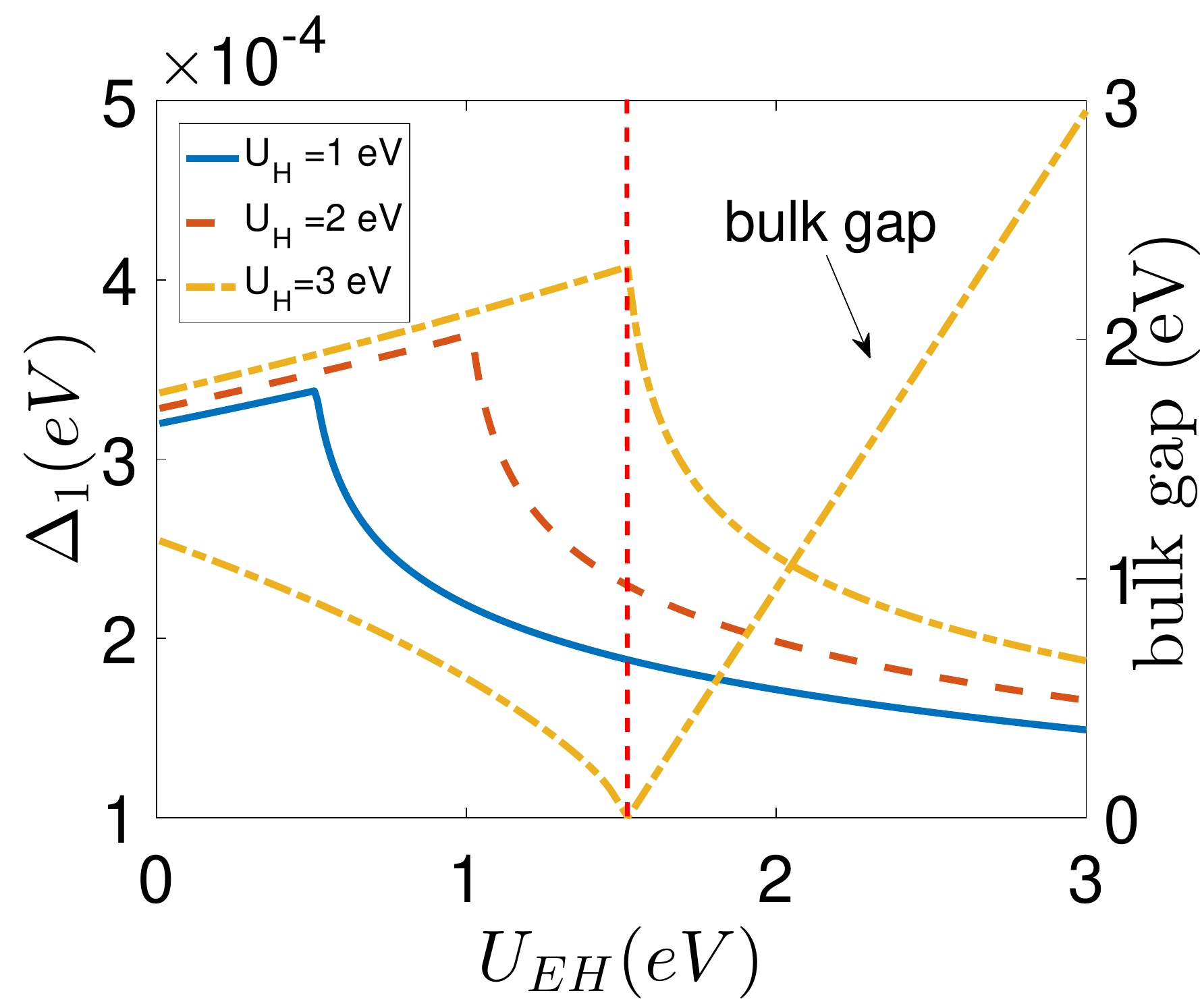}}
 \quad
\subfigure[\label{fig:deltau}]{\includegraphics[width=0.23\textwidth]{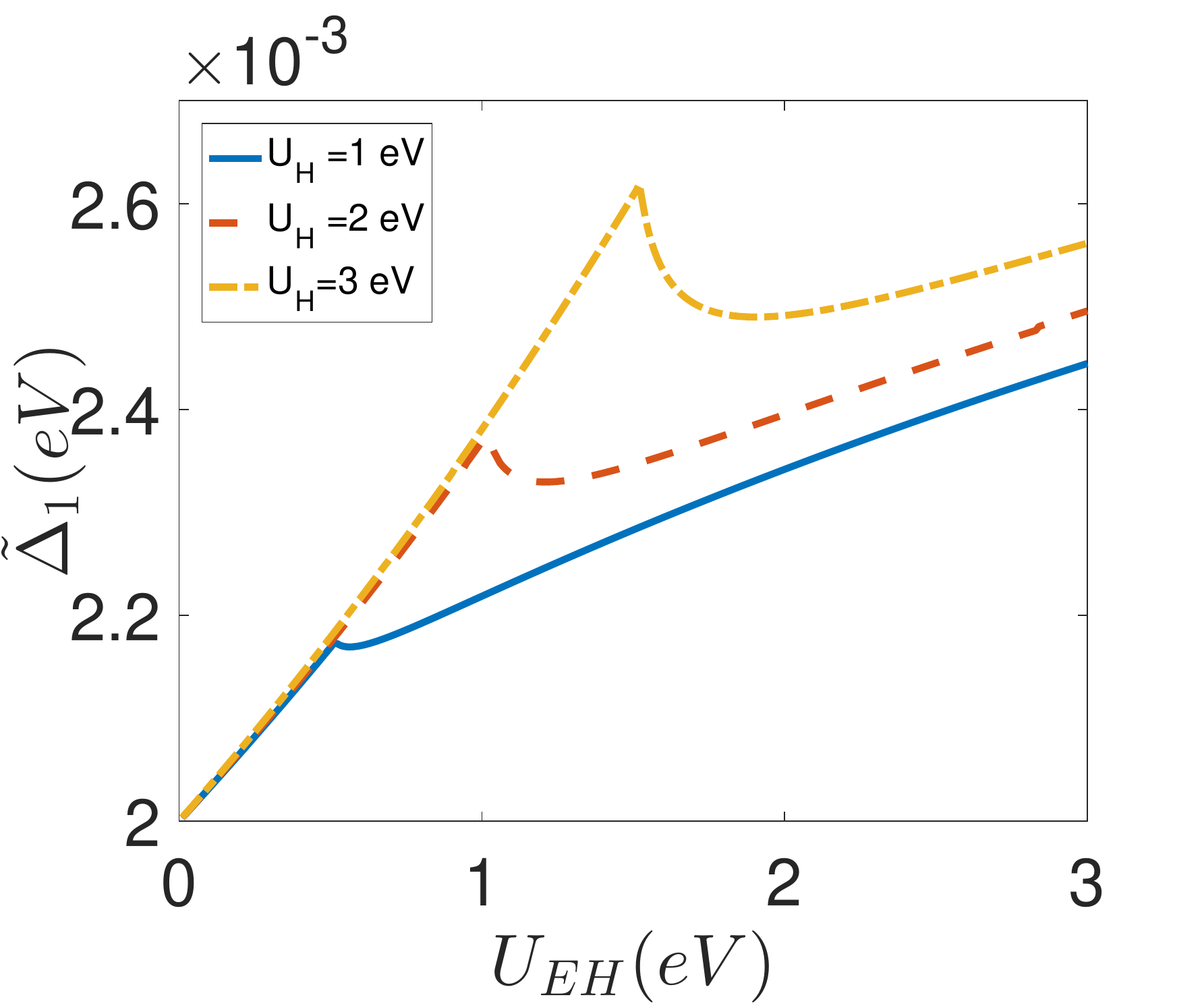}}
}
\caption{(a) Self-consistent mean-field results for $\Delta_{1}=\Delta_{2}$ as a function of $U_{EH}$ for several values of $U_{H}$ at $B_z=0$ and $U_E=1 eV$. Left axis represents $\Delta_{1(2)}$ whereas right axis represents the bulk gap. (b) Interaction modified BIA term $\tilde{\Delta}_{1}=\tilde{\Delta}_{2}$ corresponding to Fig. (1a).}
\label{Delta}
\end{figure}

 More interestingly, our mean field theory allows enhancement of BIA terms $\Delta_0\rightarrow \tilde{\Delta}_{1(2)}=\Delta_0+U_{EH}\Delta_{1(2)}$. In the following we shall consider weak interactions where the ground state is non-magnetic at $B_z=0$. In this limit $U_{E}$ has almost no effect due to the small occupation number of $E$-orbital (see Appendix B). In Fig.(\ref{Delta}), we show the calculated values of $\Delta_{1(2)}$ and the corresponding interaction modified BIA term $\tilde{\Delta}_{1(2)}$ for different interaction strengths $U_H, U_{EH}$ with fixed $U_E=1 eV$ at  $B_z=0$. We note that $\Delta_1=\Delta_2$ in this case due to TRS. We observe that $\Delta_{1(2)}$ exhibits a peak at a critical value of interaction. The peak is driven by the closing and re-opening of the bulk gap (i.e. destruction of band-inversion)  as a result of change in interaction strengths, suggesting that $\Delta_{1(2)}$ is enhanced by the resonance between the electron- and hole- energy levels. To see this we also show the bulk gap for $U_{H}=3 eV$, as a function of $U_{EH}$ in Fig.(\ref{fig:delta1}). It is clear that the peak position in $\Delta_{1(2)}$ matches with where the bulk gap closes. The interaction modified BIA term $\tilde{\Delta}_{1}=\tilde{\Delta}_{2}$ is plotted in Fig.(\ref{fig:deltau}). It gets slightly enhanced from $\Delta_0$, with a maximum enhancement of roughly 30 percent in the band closing region. The small BIA term 
   does not gap out the edge but changes the spin orientation of the helical edge states\cite{QSHETAE}. 

   Next we study the effect of magnetic field on the BIA term. We choose the interaction strengths to be $U_{E}= 5 eV, U_{H}=5 eV$ and $U_{EH}=2.5 eV$ such that the resulting mean-field band structure at zero magnetic field is almost identical to the one when all interaction strengths are set to be zero\cite{ReentrantTP}. 
   Using these parameters, we study the interaction effect on the BIA term under a perpendicular magnetic field.

\begin{figure}[tbh!]
\centering
\mbox{

\subfigure[\label{fig:LLFC}]{\includegraphics[width=0.23\textwidth]{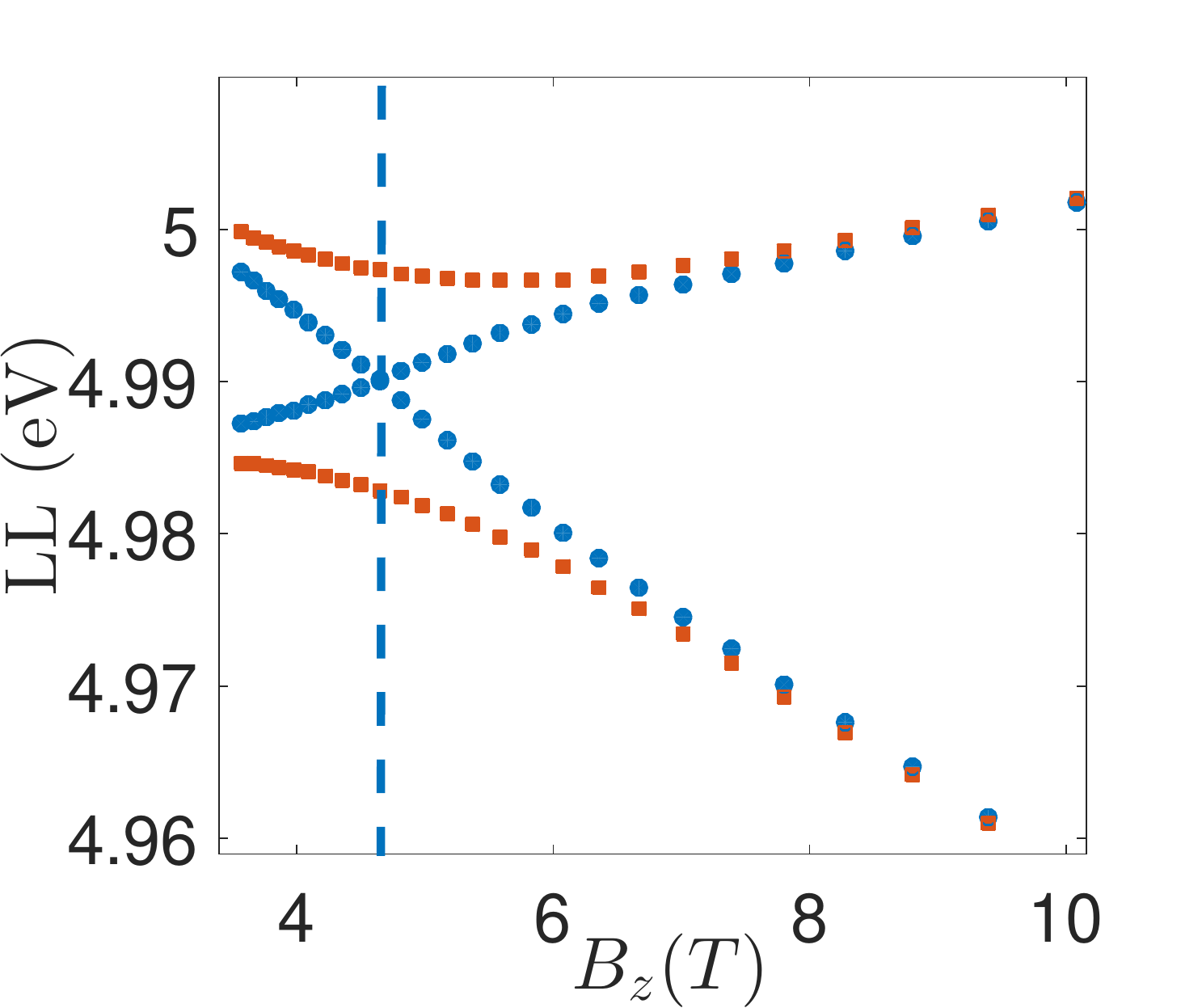}}
 \quad
\subfigure[\label{fig:delta1_Vz}]{\includegraphics[width=0.23\textwidth]{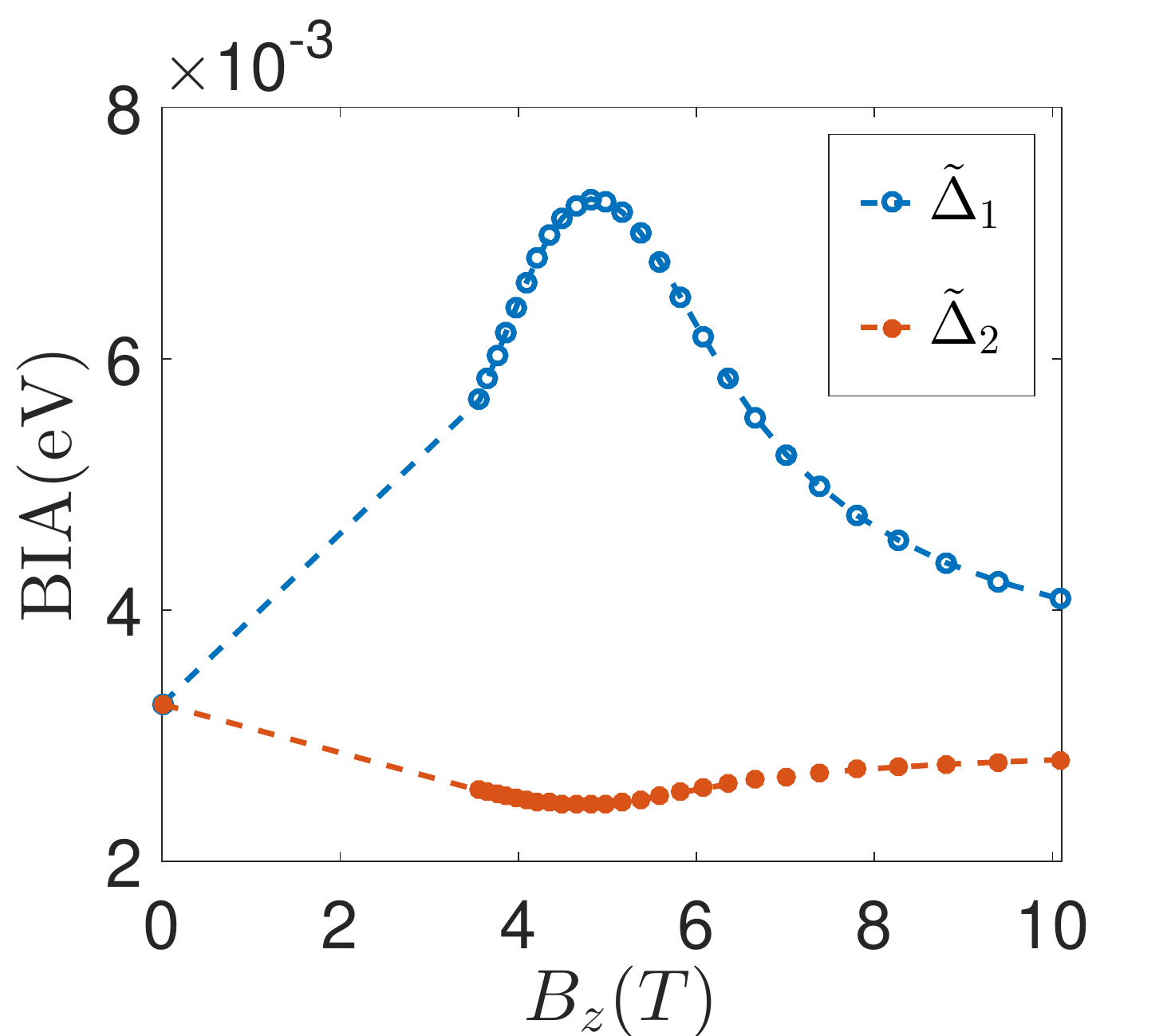}}

}
\caption{ (a) Lowest LLs as a function of $B_z$ with interaction parameters $U_{E}=5 eV, U_{H}=5 eV , U_{EH}=2.5 eV$. The dots and squares are calculated without and with the BIA term, respectively. The dash line indicates the position of the critical field.(b) Corresponding $\tilde{\Delta}_{1(2)}$ as a function a $B_{z}$ .}
\label{magnetic_field_dependence}
\end{figure}

  In Fig.(\ref{fig:LLFC}) we plot the LL without the BIA term (dots) and with the BIA term (squares). We first consider the LL without the BIA term.  In this  case the effective Hamiltonian near $\Gamma$ point reduces to two decoupled Dirac  Hamiltonian at zero magnetic field (see Appendix C). In the presence of magnetic field, LLs are formed and the zeroth LL wave function contains only one orbital component, E(H)-orbital for spin up(down). Due to the band inversion, the zeroth electron-like LL has lower energy than the zeroth hole-like LL at weak magnetic field.  As the magnetic field increases, the two zeroth Landau levels (LLs) cross at a critical magnetic field $B_c$ where the band inversion is destroyed. The system transits from a QSH state to a IQH state. The critical magnetic field is found to be around $4.5T$ which is close to the  estimation in Ref.[\onlinecite{unexpectededge}]. When the BIA term is included, the crossing of the two zeroth LLs is avoided because  of hybridization between the two LLs which is allowed when TRS is broken. In Fig.(\ref{fig:delta1_Vz}) we show the corresponding $\tilde{\Delta}_{1(2)}$ as a function of magnetic field. We note that $\tilde{\Delta}_{1}\neq\tilde{\Delta}_{2}$ in the presence of magnetic field and $\tilde{\Delta}_{1(2)}$, shows a peak(dip) at a magnetic field close to the critical magnetic field $B_c$, suggesting that the  peak(dip) in $\tilde{\Delta}_{1}\neq\tilde{\Delta}_{2}$ is driven by resonance between electron- and hole- energy levels as discussed before. This resonance is absent in trivial band-insulators where there is no band-inversion.
  The peak value of the interaction enhanced BIA term $\tilde{\Delta}_{1}$ is about 3.65 times of the bare value $\Delta_0$. On the contrary, $\tilde{\Delta}_{2}$ is only slightly enhanced but this enhancement is not important as  $\tilde{\Delta}_{1}$ is the major term responsible for the hybridization between the lowest electron and hole Landau Levels.



\begin{figure}[tbh!]
\centering
\mbox{\subfigure[\label{fig:BS_without_delta}]{\includegraphics[width=0.46\textwidth]{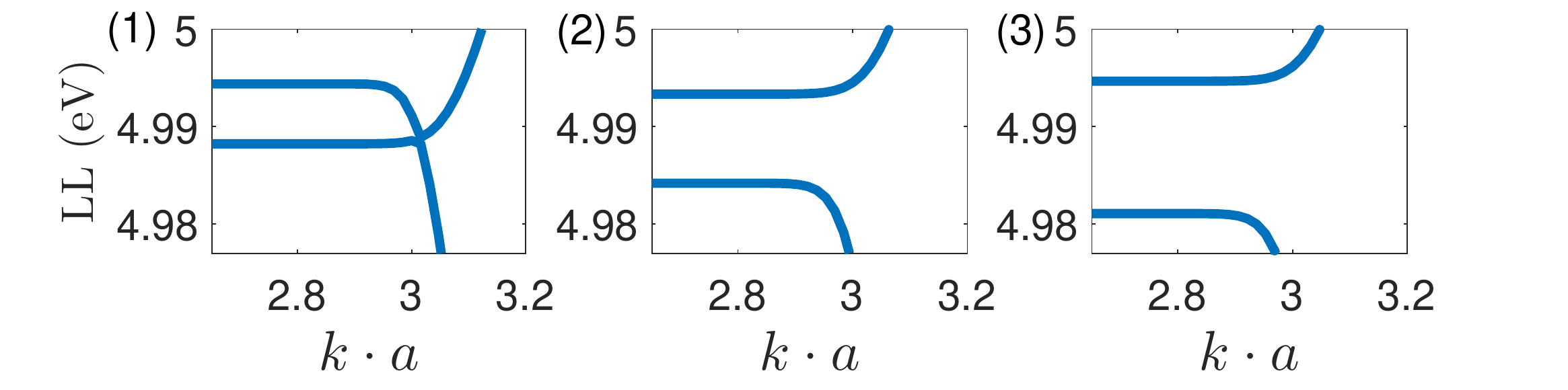}}
}
\mbox{
\subfigure[\label{fig:BS_with_delta}]{\includegraphics[width=0.46\textwidth]{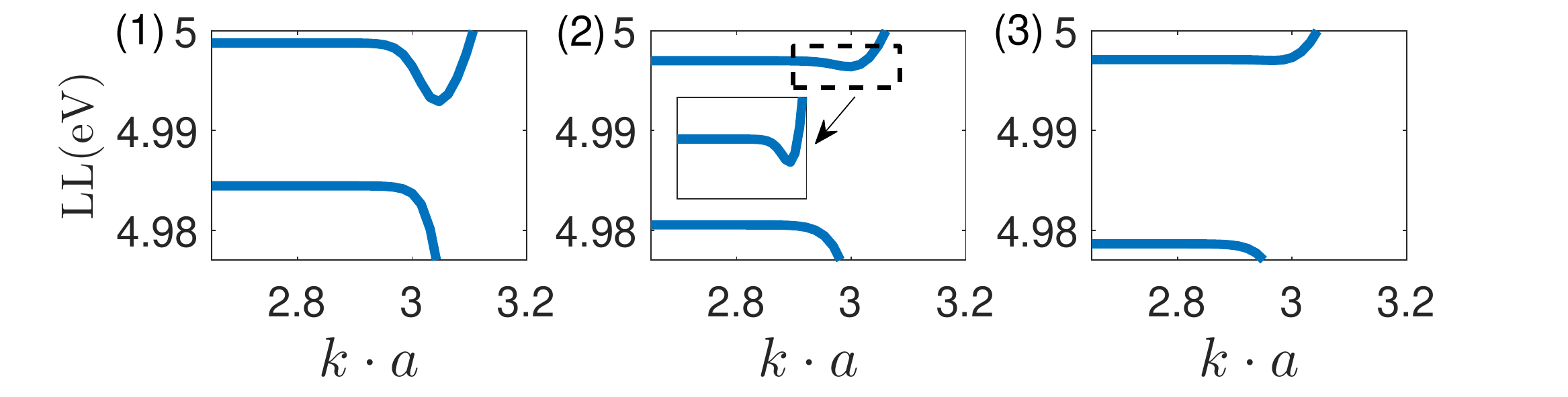}}
}

\caption{ (a) Band structure for open boundary in y-direction without  BIA term at  $U_{E}=5 eV, U_{H}=5 eV , U_{EH}=2.5 eV$ for different $B_z$. (1)-(3) is for  $B_z= 4 T, 5.5 T$ and $6 T$ respectively. (b) Band structure with BIA term  with other parameters same as in (a). The inset in (2) is a zoom-in to show the non-topological edge states. }
\label{fig:BS}
\end{figure}

 In the following we study the effect of enhanced $\tilde{\Delta}_{1}$ under magnetic field on the edge properties in our model. We consider a sample with periodic boundary in x-direction and open boundary in y-direction and calculate the corresponding band structure at different magnetic fields $B_z=4 T, 5.5 T, 6 T$, both without and with the BIA term. The result of the calculation as a function $k_x$ is shown in Fig.(\ref{fig:BS}). We show only the zeroth electron-like and hole-like LL in Fig.(\ref{fig:BS}) as they contribute to transports in Shen group's experiment\cite{unexpectededge}. Without the BIA term (Fig.(\ref{fig:BS_without_delta})),the edge is gapless when the magnetic field is smaller than the critical field $B_c\sim4.5T$. As magnetic field increases beyond $B_c$, the two zeroth LLs cross and  the system has transited from a QSH state to an IQH state. Edge transport is expected only when the zeroth LL (either electron-like or hole like) is fully filled. 

 When the BIA term is added ( Fig.(\ref{fig:BS_with_delta})),  a small gap is opened on the edge at $B_z=4 T$, but edge states with lower energies than the bulk can still be observed by slight gating. At $B_z=5.5 T$, which is beyond the critical field $B_c$, we find a small dip near the edge of the zeroth electron-like LL.  These  (non-topological) edge-like states makes the unusual edge transports beyond $B_c$ but without IQHE possible. When the system is gated,  electrons have to fill in theses edge-like states first before they occupy the bulk LL making edge conductivity possible.  We note that these edge-like states appear only in the zeroth electron-like LL but not in the zeroth hole-like LL, consistent with the experimental result that edge conductivity is observed with positive gate only. Furthermore, we also find that the BIA term decreases when the magnetic field further increases in our calculation. In particular, the non-topological edge-like states disappear and the band structure goes back to that of a normal IQH state when magnetic field is beyond a critical value $B_{e}$(see calculation result at 6T), confirming that these edge states are non-topological. We thus predict that the edge transports observed in Shen's experiment will disappear when magnetic field increases further.

How does a large BIA term create the non-topological edge-like states? 
To understand the origin of the non-topological edge-like states, we study the quantum Hall problem in a confined system with an effective low energy $\mathbf{k}\cdot \mathbf{p}$ Hamiltonian generated from our mean-field Hamiltonian $H_{MF}$ near $\mathbf{k}=0$ with the effect of edge simulated by a confining, linear orbital-dependent potential. The finding of this analysis is summarized in the following. The details of our calculation is given in Appendix C. 

 In the absence of the hybridization terms $A=2t_{EH}=\Delta_0=0$, the quantum Hall problem reduces to four decouples LLs described by harmonic oscillator Hamiltonian with eigenvalues $(n+{1\over2})\omega_{\tau} (\tau=E,H)$ and the linear potential contributes a linear k-dependent shift in energy$\sim \alpha_{\tau}(k_xl^2_B-y_0)$ (for $k_xl_B^2\gg y_0$) to the states near the edge, $n=$ Landau level index and $\l_B=\sqrt{\hbar/(e|B_z|)}$ is the magnetic length. The linear potential also shifts the wave function guiding center at the edge by an amount $\Delta y=\xi_0^{'}l_B$ where $\xi_0^{'} = 2/\pi$ (see Appendix C). $A$ and $\Delta_0$ introduce hybridization between the LLs and the Landau Level spacing is enhanced by  an hybridization gap$\sim|\Delta+cA|$, where $c\sim -\frac{\sqrt{2}A\xi_0'}{2 l_{B}}$ is nonzero only at the edge where the wave function guiding center is shifted by an amount $\Delta_y$  when $\alpha_{\tau}\neq0$ (see Appendix C). As a result, the hybridization gap is effectively reduced at the edge. This effect exists only when both $A$ and $\Delta_0$ are non-zero and competes with the linear k-dependent term which tends to increase the energy gap between the electron- and hole- LLs. When the BIA term is large enough, the later effect dominates in a narrow region of k near the edge. This leads to the appearance of non-topological edge states.

 It's interesting to note that the size of region $S_{\tau}$ where these non-topological edge appears in band-$\tau$ is found to be proportional to the band mass $\sim t_{\tau}^{-1}$ (see Appendix C). For $D>0$ ($B<0$), corresponding to $|t_{E}|<|t_{H}|$, we find $S_{E} > S_{H}$, consistent with our observation that edge-like states exist only in the electron-like LL (see Fig.(\ref{fig:BS_with_delta})) and in agreement with Shen's experiment. We note that this conclusion will be inverted if we choose $D<0$. This is why we expect that the sign of D is inverted in HgTe/Hg$_{0.3}$Cd$_{0.7}$Te .



\section{Conclusion}

Summarizing, we study in this paper the interaction effect in doped HgTe/CdTe quantum well using a Hubbard-type model. In the weak interaction regime where the system is not magnetically ordered at zero magnetic field,  we show that the BIA term is enhanced and exhibits a peak when the system undergoes a band-closing, re-opening transition, either driven by interaction or magnetic field. The BIA term is allowed because our system breaks inversion symmetry. The effect is small in zero magnetic field,  but the BIA term is enhanced dramatically when the band-closing, re-opening transition is driven by magnetic field, i.e. QSH to IQH transition. The large BIA term introduces strong hybridization between the zeroth spin up electron-like LL with zeroth spin down hole-like LL  and leads to the formation of edge-like states near the edge which may contribute to edge conductivity in low carrier density when the magnetic field is not too strong. Our result explains the 'unexpected'  particle-hole asymmetric edge conductivity found in experiment\cite{unexpectededge} and predicts that the BIA term will decrease again when the magnetic field increases further leading to vanishing of edge conductivity.

We thank RGC for support through Grant No. C6026-16W.

\appendix
\section{Tight binding parameters \label{tight_binding_parameters}}
 Here we outline how our tight binding Hamiltonian parameters are determined from Ref.[\onlinecite{ReentrantTP}].  Fourier transforming $H_{BHZ}$, we obtain
\begin{align}
\begin{split}
&H_{BHZ}=\sum_{\mathbf{k}} \Psi^{\dag}_{\mathbf{k}}\left(
\begin{array}{cc}
 h(\mathbf{k}) & 0 \\
 0 & h^*(\mathbf{-k})\\
\end{array}
\right)\Psi_{\mathbf{k}} \label{Hk} \\
&h(\mathbf{k})=\varepsilon_\mathbf{k}I_2+d_{\alpha}(\mathbf{k})\cdot \sigma^{\alpha}
\end{split}
\end{align}
 where $\Psi_{\mathbf{k}}=\{ C_{E,\mathbf{k},\uparrow}, C_{H,\mathbf{k},\uparrow},C_{E,\mathbf{k},\downarrow}, C_{H,\mathbf{k},\downarrow} \}^T$, $\sigma^{\alpha}$'s are Pauli matrices,
\begin{align}
C_{\tau,\mathbf{k},\sigma}=\frac{1}{\sqrt{N}} \sum_{i}\exp(i\mathbf{k} \cdot \mathbf{R}_{i}) C_{i,\tau,\sigma}
\end{align}
where N is the total number of sites and
\begin{align}
\begin{split}
&\varepsilon_{\mathbf{k}}=C-\frac{2D}{a^2}(2-\cos(k_x)-\cos(k_y)) \\
&d_{\alpha}(\mathbf{k})=[\frac{A}{a}\sin(k_x),-\frac{A}{a}\sin(k_y),M(\mathbf{k})]\\
&M(\mathbf{k})=M-\frac{2B}{a^2}(2-\cos(k_x)-\cos(k_y)) \\
D&=(t_E+t_H)/2, B=(t_E-t_H)/2, A=2t_{EH}\\
M&=\frac{\varepsilon_E-\varepsilon_H}{2}-2(t_E-t_H)\\
C&=\frac{\varepsilon_E+\varepsilon_H}{2}-2(t_E+t_H)
\end{split}
\end{align}

Expanding Eq.(\ref{Hk}) around $\mathbf{k}=0$ we obtained the Hamiltonian (1) in Ref.[\onlinecite{ReentrantTP}]. All the tight binding parameters and the $g_{\tau}$ factors can be identified from Table.1 of Ref.[\onlinecite{ReentrantTP}].

\section{Mean field phase diagram}
 We discuss the effect of interaction on HgTe/CdTe quantum well at zero magnetic field in this appendix. The mean field Hamiltonian is:
\begin{align}
\begin{split}
H_{MF}&=H_{BHZ}+ H_{BIA}+ H_{z}
\\&+\sum_{i,\sigma,\tau}(U_{\tau} \langle n_{i,\tau,-\sigma} \rangle +U_{EH} \langle n_{i,\bar{\tau}} \rangle) n_{i,\tau,\sigma}\\
&-U_{EH}(\Delta_1 C_{i,E,\uparrow}^{\dag}C_{i,H,\downarrow}-\Delta_{2} C_{i,H,\uparrow}^{\dag}C_{i,E,\downarrow}+ \text{H.c})
\end{split}
\end{align}
where $\Delta_{1(2)}=(-)\langle C_{i,H(E),\downarrow(\uparrow)}^{\dag}C_{i,E(H),\uparrow(\downarrow)}\rangle$ and $n_{i,\tau}=\sum_{\sigma}n_{i,\tau,\sigma}$.
We note that $\Delta_1=\Delta_2$ in the absence of magnetic field.

\begin{figure}[tbh!]
\centering
\mbox{
\subfigure[\label{schematicband1}]{\includegraphics[width=0.23\textwidth]{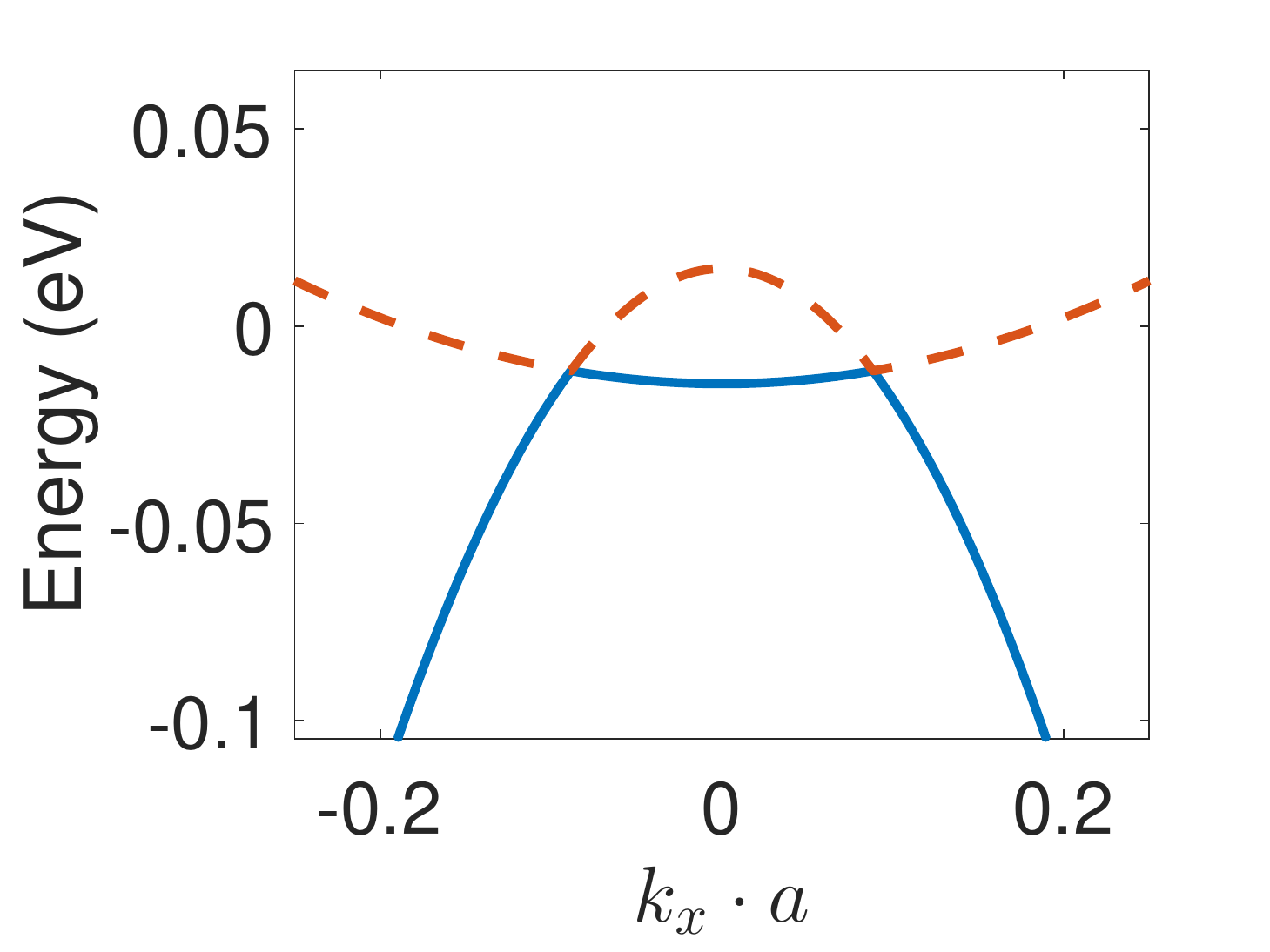}}
 \quad
\subfigure[\label{schematicband2}]{\includegraphics[width=0.23\textwidth]{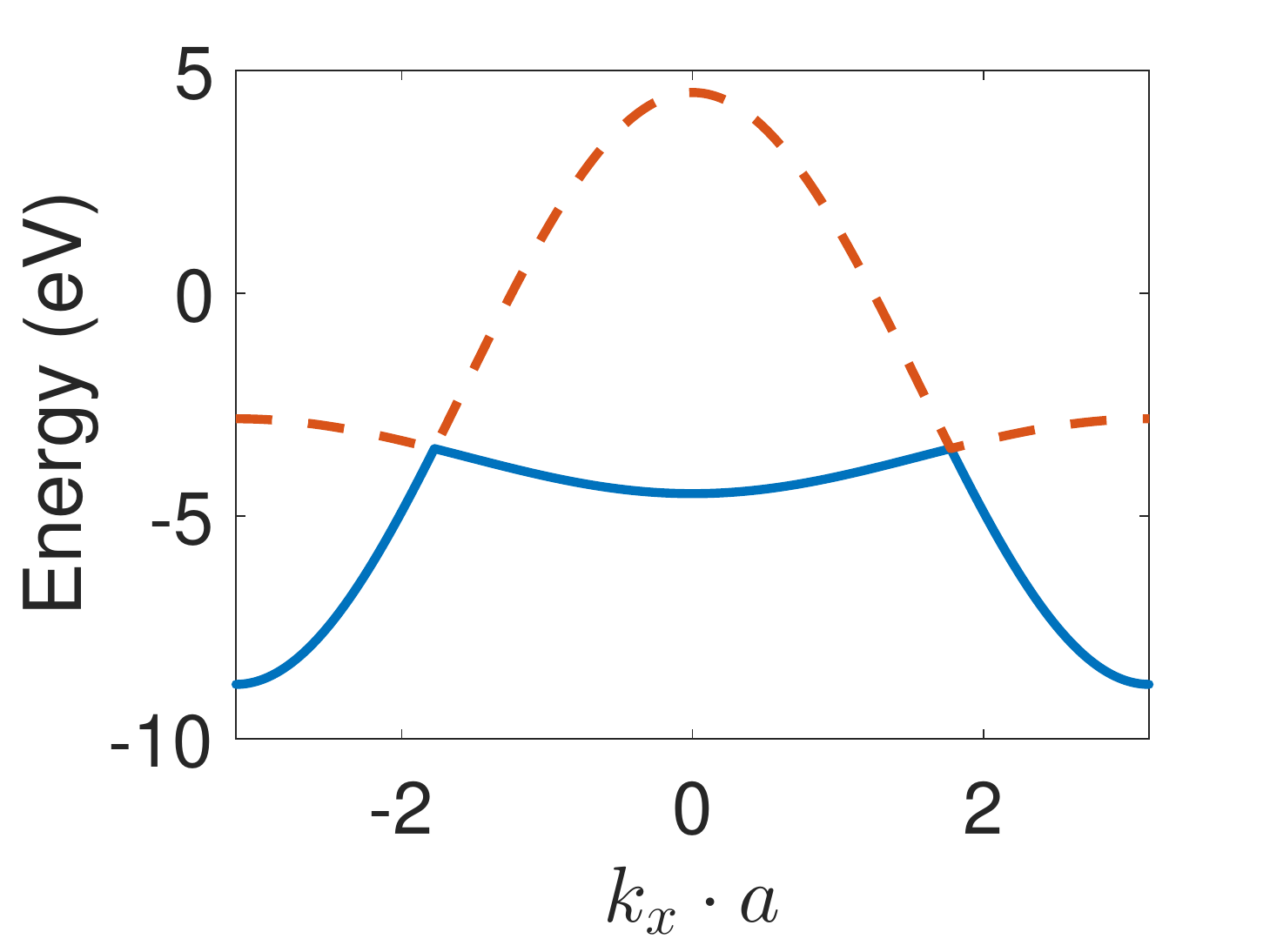}}
}
\caption{Schematic band structure illustrating the filling of the E and H bands. The solid/dash lines denotes the part of bands which are occupied/empty. (a) The case for small band inversion corresponding to HgTe/CdTe quantum well. (b) The situation with large band inversion. }.
\label{schematicband}
\end{figure}

To understand the physics behind the mean-field results, we first consider the case when the hybridization between the E and H orbital ($t_{EH}$ and $\Delta_0$) vanishes.  In this case, the $E$- and $H$- orbital form separate bands which overlap because of band inversion (see Fig.(\ref{schematicband})). A small part of the E-band is occupied whereas the H-band is almost filled (see Fig.(\ref{schematicband1})). In this case, the E and H bands are described separately by single-band Hubbard models which are almost empty/filled. Mean-field studies for single-band Hubbard model on square lattice has been carried out long time ago\cite{PhysRevB.31.4403} and it was found that the ground state is {\em anti-ferromagnetic} at and close to half filling and becomes {\em ferromagnetic} away from half filling when the interaction strength $U$ is large than certain critical value.  In HgTe/CdTe quantum well the E and H bands are nearly empty or fully filled at weak interaction limit suggesting that we should look for ferromagnetic phases in our mean-field theory. Anti-ferromagnetic phase is expected only if the band inversion is so large that the two bands are both nearly half filled (case shown in Fig.(\ref{schematicband2})).
We search for the {\em paramagnetic , ferromagnetic} and {\em anti-ferromagnetic} phases numerically in our study starting from the half filled case for the BHZ model where the chemical potential is in the gap and the system is a topological insulator. We employ the parameters  as discussed in the main text where $t_{E}=-0.42 eV, t_{H}=3.32 eV, t_{EH}=0.275 eV,\varepsilon_{E}=1.67 eV, \varepsilon_{H}=-13.27 eV$ \cite{ReentrantTP}. $\Delta_0=0.002 eV$ The lattice constant $a$ is chosen to be 1 nm.

\begin{figure}[tbh!]
\centering
\mbox{
\subfigure[\label{schematicband1}]{\includegraphics[width=0.23\textwidth]{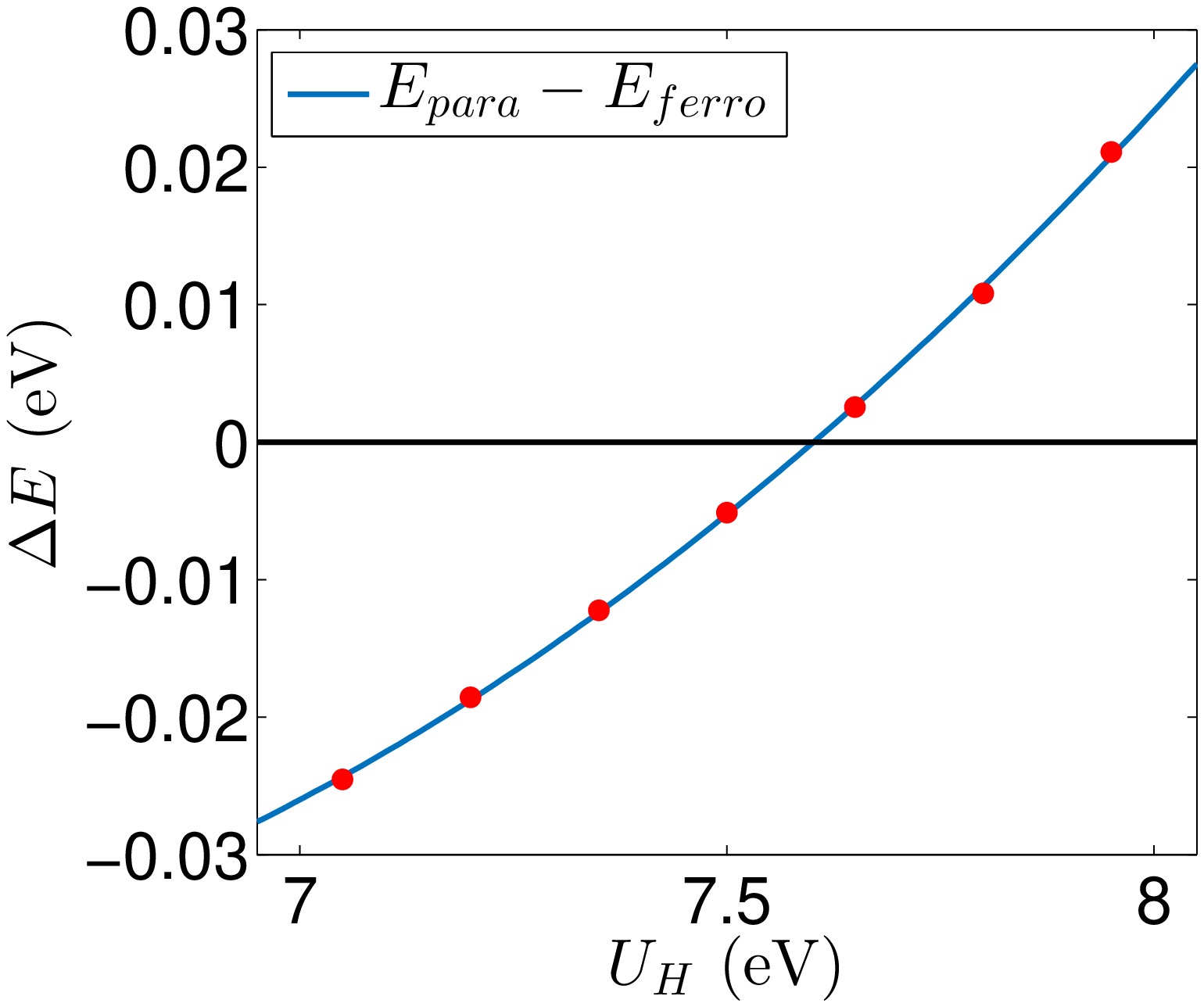}}
 \quad
\subfigure[\label{schematicband2}]{\includegraphics[width=0.23\textwidth]{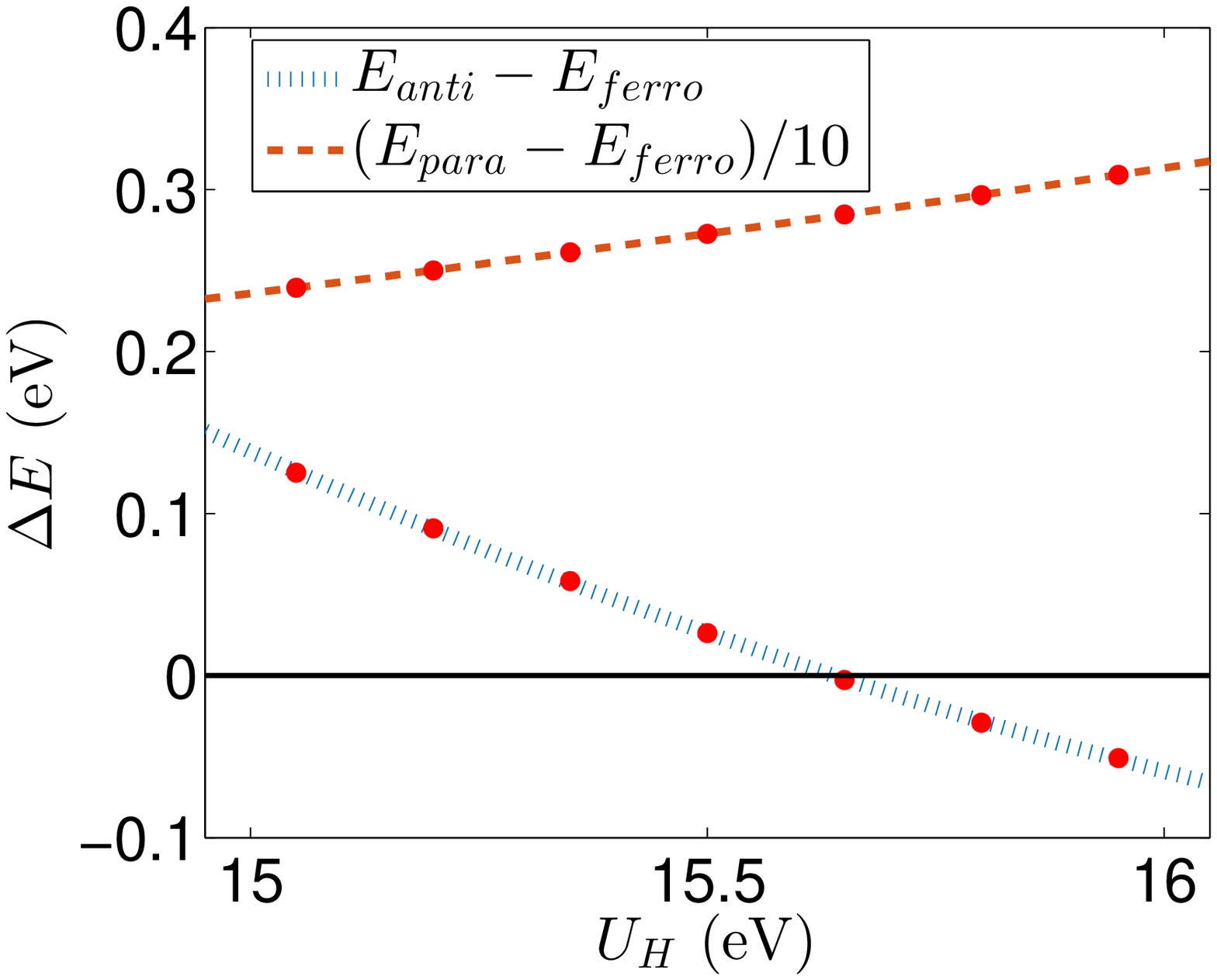}}
}
\caption{Energy difference ($\Delta_E$) between different phases as a function of $U_{H}$ with $U_{E} = 0, U_{EH} = 1 eV$ .  The dots are data calculated self consistently. (a) Energy difference between {\em paramagnetic} phase and {\em ferromagnetic phase} at small $U_H<10$eV. There is no {\em stable anti-ferromagnetic} phase found in this region. (b)  Energy difference between {\em paramagnetic} phase and {\em ferromagnetic phase} and  Energy difference between {\em anti-ferromagnetic} phase and {\em ferromagnetic phase}  at large $U_H>15$ eV. We note that the {\em anti-ferromagnetic} phase becomes the ground state only at very large $U_H$ }. 
\label{enengy_difference}
\end{figure}

We first consider the case with only $U_H\neq0$ which is similar to the single band Hubbard model. We note that an important difference between the single band Hubbard model and the BHZ model is that in our case, the relative position of the two bands depends on interaction.  When $U_H$ increases, the on-site energy of H orbital is shifted upward while the E orbital energy remains stationary leading to increasing population in E band. Changing other interactions have similar effects. Thus we are actually moving along a curve in the density-interaction phase diagram of  an effective one-band Hubbard model when interaction changes.   For small $U_{H}$, only one solution with $m_{H}=m_{E}=0$ is found.  As interaction strength increase, two self-consistent solutions appear. The ground state is the one with lower energy. For illustration, we shown the energy difference between different phases as a function of $U_H$ with $U_E=0$, $U_{EH}=1eV$ in Fig.(\ref{enengy_difference})

\begin{figure}[tbh!]
\centering
\mbox{
\subfigure[]{\includegraphics[width=0.23\textwidth]{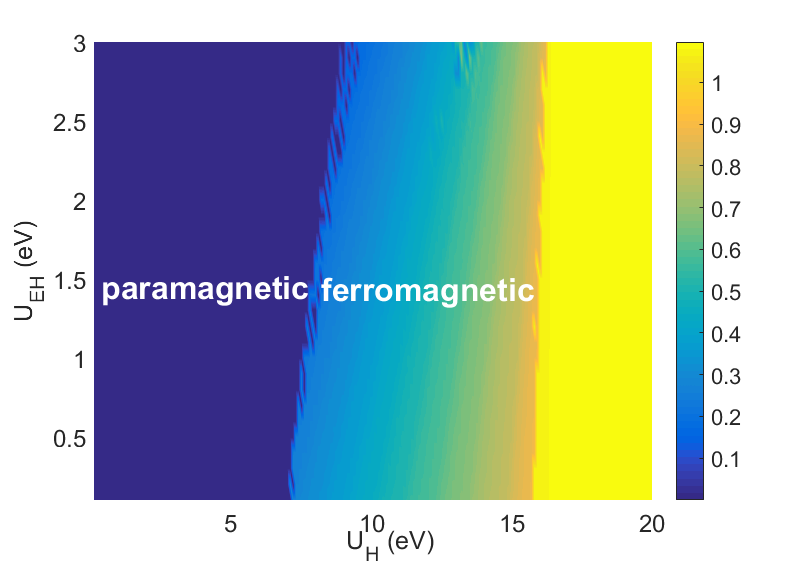}}
 \quad
\subfigure[]{\includegraphics[width=0.23\textwidth]{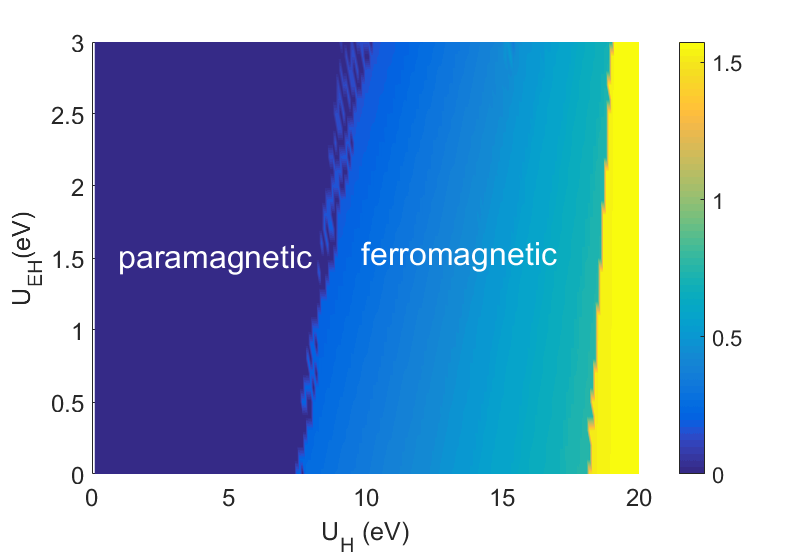}}
}
\caption{The mean field phase diagram as a function of $U_{H}$ and $U_{EH}$ for (a) $U_{E}=0$ (b) $U_{E}=10 eV$. Left region : paramagnetic phase, middle region: ferromagnetic phase , right region: anti-ferromagnetic phase. The color indicates the magnitude of the magnetic order.}
\label{phase_diagram}
\end{figure}

Including $U_E$ and $U_{EH}$ have the similar effect as $U_H$. $U_E$ increases the energy of E orbital. However, as discussed above, when $U_H$ is weak the occupation number of E orbital is much smaller than H, and the effect of $U_E$ is much smaller compared to $U_H$ because of the smallness of $n_E$.  Therefore $U_E$ has almost no effect on the phase transition in the weak $U_{H}$ limit.  $U_{EH}$ raises the energies of the two orbital simultaneously but with different values depending on the occupation numbers of the two bands. The shift in the energy of E(H) orbital is proportional to $n_H(n_E)$. Again, since $n_H>> n_E$, the energy of E orbital is shifted faster than H orbital leading to decreasing/increasing occupation number in E/H orbital for $U_{EH}>0$. The role of $U_{E}$ and $U_{EH}$ reverses in the large $U_{H}$ limit when  $n_E$ becomes comparable to $n_{H}$. 

 The dependence of the {\em paramagnetic-ferromagnetic-anti-ferromagnetic} phase boundary on the interactions are summarized in the phase diagram in Fig.(\ref{phase_diagram}).  Comparing the two phase diagrams for $U_E=0$ and $U_E=10 eV$, we see that a large $U_E$ shifts the {\em paramagnetic-ferromagnetic} boundary only slightly, but it shifts the {\em ferromagnetic-anti-ferromagnetic} boundary more significantly, in agreement with our analysis.

\section{Hybridization between Landau levels and edge-like states}
  We discuss here how hybridization between electron- and hole- like Landau levels leads to the emergence of the edge-like states. We start with considering the quantum Hall problem using an effective $\mathbf{k}\cdot \mathbf{p}$ Hamiltonian generated from our mean-field Hamiltonian $H_{MF}$ near $\mathbf{k}=0$ with the effect of edge simulated by a confining, linear orbital-dependent potential $V_{\tau}(y)$ (we assume here the edge is along x-direction). For simplicity we neglect the Zeeman energy term and assume $\tilde{\Delta}_1=\tilde{\Delta}_2=\Delta$ in our following calculation. The effective Hamiltonian is thus\cite{RevModPhys.83.1057},
 \begin{subequations}
 \begin{equation}
 \label{effH}
 H=\sum_{\mathbf{k}}\bar{\Psi}_{\mathbf{k}}\mathbf{H}_{kp}(\mathbf{k})\Psi_{\mathbf{k}}+\int d^2r V_c(y)\bar{\Psi}(\mathbf{r})\Psi(\mathbf{r}),
 \end{equation}
 where $\Psi(\mathbf{r})=\{\Psi_{E,\uparrow}(\mathbf{r}),\Psi_{H,\uparrow}(\mathbf{r}),\Psi_{E,\downarrow}(\mathbf{r}),\Psi_{H,\downarrow}(\mathbf{r})\}^{T}$ and $\Psi_{\mathbf{k}}$ is the Fourier transform of $\Psi(\mathbf{r})$.
\begin{align}
\mathbf{H}_{kp}&= -\sigma_0\tau_0 D k^2 + \sigma_0 \tau_z(M-B k^2a^2) \nonumber \\
&+\sigma_z \tau_x A k_x - \sigma_0\tau_y A k_y+\sigma_y\tau_y \Delta
\end{align}
where $\sigma_i,\tau_i,i=x,y,z$ are Pauli matrix acting on spin basis and orbit basis respectively. $\sigma_0,\tau_0$ is the corresponding $2 \times 2$ identity matrix. $k^2=
k_x^2+k_y^2$. $D=(t_{E}+t_{H})a^2/2,B=(t_{E}-t_{H})a^2/2,M=\varepsilon_{E}+4t_{E},A=2t_{EH}a$.
\begin{equation}
V_{\tau}(y)=
\begin{array}{c}
-\alpha_{\tau}(y+y_0), ~~~y<-y_0\\
~~~~~~~0, ~~~~~~~-y_0<y<y_0\\
\alpha_{\tau}(y-y_0), ~~~y>y_0
\end{array}
\end{equation}
\end{subequations}
is a linear confining potential at the edge which vanishes in the bulk. $\alpha_E>0$ and $\alpha_H <0$ for the electron- and hole- like orbital, respectively. We consider the Landau gauge $\mathbf{A}=-B_z y \hat{x}$ such that $H$ is translational invariant along $x$-direction and $k_x$ is a good quantum number. In this case, we may replace $k_y$ by $k_y\to -i\hbar \partial_y-e B_z y$ and $H$ becomes,
\begin{align}
H\rightarrow \left(
\begin{array}{cccc}
 K_E & \eta  f^{\dagger} & 0 & -\Delta  \\
 \eta  f &  K_H & \Delta  & 0 \\
 0 & \Delta  & K_E & -\eta  f \\
 -\Delta  & 0 & -\eta  f^{\dagger} & K_H \\
\end{array}
\right)
\end{align}
where $f=\frac{\xi}{2}+\partial_{\xi} ,f^{\dagger}=\frac{\xi}{2}-\partial_{\xi}$ and $\xi=\sqrt{2}(y-l_B^2 k_x)/l_B, \eta=-\sqrt{2}A/l_B$. $K_{\tau}=M_{\tau}+\omega_{\tau}\left(f^{\dagger} f+\frac{1}{2}\right)+V_{\tau}(y)$  where $\omega_{\tau}=-2t_{\tau}a^2/l_B^2$ and $M_{\tau}=\varepsilon_{\tau}+4t_{\tau}$. $K_{\tau}$ is the usual Harmonic Oscillator type Hamiltonian describing electrons/holes moving in a single orbital and the rest of the terms describe hybridization between different orbital.

To show how edge-like states emerge we assume that $A$ and $\Delta$ are small compare with Landau level spacings and treat them as perturbations.
First we consider $A=0,\Delta=0$. In this case the eigenvalues and wave functions at the right edge ($k_xl_B^2>y_0$) are given by,
\begin{align}
\varepsilon_{n,\sigma}^{\tau}(k_x)&=M_{\tau}+\omega_{\tau}(n+\frac{1}{2})+\alpha_{\tau}(k_x l_{B}^2-y_0)\\
\langle\xi|\phi_{n,\tau,\sigma}(k_x)\rangle&=\frac{\exp(i k_x x)}{\sqrt{N}}\frac{\exp(-(\xi-\xi_n')^2/2)}{\sqrt{2^n n! \sqrt{\pi}}} H_n(\xi-\xi_n')v_{\tau,\sigma} \label{wavefunction}
\end{align}
where $n$'s are Landau level indices, $H_n$ is the hermitian polynomial, $v_{E,\uparrow}=\{1,0,0,0\}^T, v_{H,\uparrow}=\{0,1,0,0\}^T, v_{E,\downarrow}=\{0,0,1,0\}^T,v_{H,\downarrow}=\{0,0,0,1\}^T$. N is the number of sites in x direction. The first two terms in $\varepsilon_{n,\sigma}^{\tau}(k_x)$ describe the bulk LL energy. The last term, which is linear in $k_x$, is the result of the linear  edge potential. Besides the linear dispersion, the linear potential also shifts the wave function guiding center by an amount $\xi_n' = \alpha_{\tau}\frac{\sqrt{2}l_{B}^3}{2t_{\tau}a^2}$ in Eq.(\ref{wavefunction}).

To determine the value of $\alpha_{\tau}$, we notice that the linear potential gives rise to a drift velocity $v_{d,\tau}=\alpha_{\tau}l_{B}^2/\hbar$ along the edge. The slope $\alpha_{\tau}$ can be determined by comparing this drift velocity with the drift velocity computed for IQH states with sharp edge, where the semi-classical picture gives $v_{d,\tau}=2/\pi \sqrt{\omega_{\tau}(n+1/2)/m_{\tau}^{*}}$. $m_{\tau}^* \approx -\hbar^2 /(t_{\tau} a^2) $ is the effective mass of $\tau$ orbital near the band edge. Comparing the two results, we find that $\alpha_{\tau} \sim -\frac{t_{\tau}a^2}{l_B^3 }$. Substituting into $\xi_n'$ we find that the wave function shift depends on the LL index only, with $\xi_{n}'=2\sqrt{2}/\pi \sqrt{n+1/2}$.

 When $A$ and $\Delta$ is turned on,  $A$ couples in the bulk $|\phi_{1,E,\uparrow}(k_x)\rangle $ with $|\phi_{0,H,\uparrow}(k_x)\rangle$ and $|\phi_{0,E,\downarrow}(k_x)\rangle$ with $|\phi_{1,H,\downarrow}(k_x) \rangle$. What is interesting is that the $n=0$ electron and hole levels  $|\phi_{0,E,\sigma}(k_x) \rangle$ and $|\phi_{0,H,\sigma}(k_x) \rangle$ are also coupled at the edge due to the shift in the guiding center of the wave functions. With this in mind we write down an effective  Hamiltonian for the $n=0$ LLs.  In the basis, $\{|\phi_{0,E,\uparrow}(k_x) \rangle, |\phi_{0,H,\uparrow}(k_x) \rangle, |\phi_{0,E,\downarrow}(k_x)\rangle, |\phi_{0,H,\downarrow}(k_x)\rangle\}^{T}$ the effective Hamiltonian becomes,

\begin{align}
H_{0} &= \left(
\begin{array}{cc}
 H_{0,\uparrow} & H_{\Delta} \\
 H_{\Delta}^{\dagger} & H_{0,\downarrow} \\
\end{array}
\right)
\label{H00}
\end{align}
\begin{align*}
H_{0,\sigma}=\left(
\begin{array}{cc}
 \varepsilon_{0,\sigma}^{E} & s\eta h_0 \\
 s\eta h_0 & \varepsilon_{0,\sigma}^{H} \\
\end{array}
\right),
 &H_{\Delta}=\left(
\begin{array}{cc}
 0 & -\Delta \\
 \Delta & 0 \\
\end{array}
\right)
\end{align*}
where $s=+(-)1$ for $\sigma=\uparrow(\downarrow)$ and
\begin{align*}
h_0&=\langle \phi_{0,E,\uparrow}(k_x) |f^{\dagger}\sigma_z\tau_x|\phi_{0,H,\uparrow}(k_x)\rangle \numberthis \\
&=\langle \phi_{0,E,\uparrow}(k_x)|\frac{\xi-\xi_0'}{2}-\partial_{\xi-\xi_0'} +\xi_0'/2 |\phi_{0,E,\uparrow}(k_x) \rangle\\
&=\xi_0'/2
\end{align*}
is the matrix element describing the (same spin) electron-hole hybridization. $\sigma_z\tau_x$  is the operator that flips the orbital index. $h_0$ vanishes in the bulk and is non-zero only in the edge due to the shift in the wave function guiding center by the linear potential as illustrated above. It's straightforward to diagonalize $H_0$ to obtain the eigen-energies
\begin{subequations}
\begin{align}
\varepsilon_{p,\pm}(k_x) &= \frac{\varepsilon_{0,\uparrow}^{E}(k_x)+\varepsilon_{0,\downarrow}^{H}(k_x)}{2} \pm \varepsilon_{p,0}(k_x),
\end{align}
where
\begin{align}
\varepsilon_{p,0}(k_x)&=\sqrt{\left(\frac{\varepsilon_{0,\uparrow}^{E}(k_x)-\varepsilon_{0,\downarrow}^{H}(k_x)}{2}\right)^2 +(\Delta+\eta\xi_0'/2)^2}
\end{align}
and
\begin{align}
\varepsilon_{m,\pm}(k_x) &= \frac{\varepsilon_{0,\downarrow}^{E}(k_x)+\varepsilon_{0,\uparrow}^{H}(k_x)}{2} \pm \varepsilon_{m,0}(k_x),
\end{align}
where
\begin{align}
\varepsilon_{m,0}(k_x)&=\sqrt{\left(\frac{\varepsilon_{0,\downarrow}^{E}(k_x)-\varepsilon_{0,\uparrow}^{H}(k_x)}{2}\right)^2 +(\Delta-\eta\xi_0'/2)^2}.
\end{align}
\end{subequations}

\begin{figure}[tbh!]
\centering
\mbox{
{\includegraphics[width=0.4\textwidth]{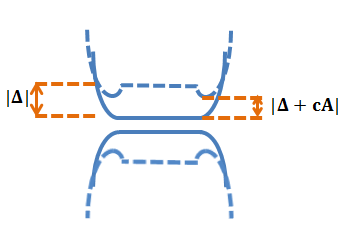}}
}

\caption{Schematic band structure to illustrate the emergence of edge-like state for the zeroth LL. Without BIA term, the zeroth LL has same spin electron-hole hybridization only at the edge(solid line) arising from $A$-term $\sim t_{EH}$. When BIA term is turn on, it opens a hybridization gap (dashed line) at the bulk while the effective hybridization $\sim |\Delta+cA|$ is weakened at the edge.}
\label{Sketch_band_structure}
\end{figure}

The first term under the square-root is the unperturbed LL spacing. The second term, $(\Delta\pm \eta \xi_0'/2)^2$ is the  hybridization contributed by $A $ and $\Delta$.  The low energy sector is described by $\varepsilon_{p,\pm}$ since $\eta<0$ from our band parameters. We notice that the hybridization term $(\Delta +\eta \xi_0'/2)$ at the edge is smaller than that of in the bulk ($\sim\Delta$) as long as $\Delta > -\eta \xi_0' /4$.  This effect exists only when both $\eta\sim A$ and $\Delta$ are nonzero.  This physical picture is illustrated in Fig.(\ref{Sketch_band_structure}).

As a result, it is possible that $\epsilon_{p,+(-)}(k_x)$ at edge ($|k_x|l_B^2>y_0$) is smaller than their value at bulk ($|k_x|l_B^2<y_0$). Assuming $\Delta_{0}+\eta \xi_0'/2\ll (\varepsilon_{0,\uparrow}^{E}(k_x)-\varepsilon_{0,\downarrow}^{H}(k_x))/2$, we obtain
\[
\epsilon_{p,+(-)}(|k_x|<y_0)\sim\omega_{E(H)}\pm{\Delta^2\over\omega_E-\omega_H}, \]
and
\begin{align}
\epsilon_{p,+(-)}(|k_x|>y_0)\sim\omega_{E(H)}+\alpha_{E(H)}(|k_x|l_B^2-y_0)\\
\pm{(\Delta+\eta\xi_0'/2)^2\over\omega_E-\omega_H+(\alpha_E-\alpha_H)(|k_x|^2l_B^2-y_0)}.
\end{align}
We have neglected $M_\tau$ since its not important beyond critical magnetic field $B_c$ defined in main text. It is easy to see that there exists a finite region $y_0<k_x l_B^{2}<k_c l_{B}^2$ where $\epsilon_{p,+(-)}(|k_x|l_{B}^2<y_0)>\epsilon_{p,+(-)}(|k_x|l_{B}^2>y_0)$.

 By keeping terms up to first order in $k_x l_B^2-y_0$, $k_c$ is given by,
\begin{equation}
\Delta^{\tau}+\frac{\xi_0'\eta}{4}\approx  -\frac{2(t_{H}-t_{E})|t_{\tau}|a^4(k_c l_B^2-y_0)}{\eta \xi_0' l_B^{5}}.
\end{equation}

 We note that $k_c$ depends on the magnitude of hopping $t_{\tau}$. When $|t_{E}|<|t_{H}|$, corresponding to $D>0$, $k_c^{E}>k_c^{H}$ and the non-topological edge state is easier to observe in the electron-like LL, consistent with the experimental result. Therefore, we chose $D>0$ for the calculation in the main text.
 
\bibliographystyle{h-physrev2}
\bibliography{ref}
\end{document}